\documentclass{PoS}

\title{The systematic spectral analysis of radio surveys}

\ShortTitle{The systematic spectral analysis of radio surveys}

\author{\speaker{Jeremy J. Harwood}\\
        ASTRON, The Netherlands Institute for Radio Astronomy, Postbus 2, 7990 AA, Dwingeloo, The Netherlands\\
        E-mail: \email{Jeremy.Harwood@physics.org}}

\author{Raffaellla Morganti\\
        ASTRON, The Netherlands Institute for Radio Astronomy, Postbus 2, 7990 AA, Dwingeloo, The Netherlands\\}

\abstract{Current and future continuum surveys being undertaken by the new generation of radio telescopes are now poised to address many important science questions, ranging from the earliest galaxies, to the physics of nearby AGN, as well as potentially providing new and unexpected discoveries. However, how to efficiently analyse the large quantities of data collected by these studies in order to maximise their scientific output remains an open question. In these proceedings we present details of the surveys module for the Broadband Radio Astronomy Tools (\textsc{brats}) software package which will combine new observations with existing multi-frequency data in order to automatically analyse and select sources based on their spectrum. We show how these methods can been applied to investigate objects observed on a variety of spatial scales, and suggest a pathway for how this can be used in the wider context of surveys and large samples.}

\FullConference{EXTRA-RADSUR2015 (*)\\
		20--23 October 2015\\
		Bologna, Italy

                \bigskip
                \hrule
                \bigskip

                \textnormal{(*) This conference has been organized
                  with the support of the Ministry of Foreign Affairs
                  and International Cooperation, Directorate General
                  for the Country Promotion (Bilateral Grant Agreement
                  ZA14GR02 - Mapping the Universe on the Pathway to
                  SKA)}
}

\begin{document}

\section{Introduction}

The ability of modern radio telescopes to undertake high sensitivity, broad-bandwidth surveys has raised an important question: how do we systematically characterize and investigate sources and populations in these vast volumes of data? Past surveys have led to the discovery of a range of unusual and previously unexplored objects, such as giant and remnant radio galaxies \cite{shulevski15, brienza16}, but so far these have usually been found serendipitously during investigations of nearby objects and the analysis of their spectrum has remained a largely manual process. As the observations required to discover such objects become more common, a more systematic method is required for both the discovery and analysis of these sources. This is key for not only optimising the scientific output of individual objects, but also for the study of source populations where the volume of data produced means that a manual analysis is not viable.

The Broadband Radio Astronomy Tools software package (\textsc{brats}\footnote{http://www.askanastronomer.co.uk/brats}) already provides a range of tools for the analysis of radio sources and has successfully been applied to various source types including radio galaxies \cite{harwood13, heesen14, harwood15}, cluster relics \cite{stroe14, degasperin15} and a non-thermal superbubble \cite{heesen15}. However, these studies have mainly been limited to well resolved, high fidelity, targeted observations where matched data are available over a wide frequency range, which is unlikely to be the case for both current and future surveys. Development of a surveys module for \textsc{brats} is therefore currently underway with the aim of providing a simple to use, direct route from source selection through to a detailed spectral analysis which can be applied to both surveys and large samples of radio sources (e.g. wide-field images from low-frequency observations). In the following sections we will discuss how the issues of analysing such large quantities of data will be resolved and how it will be applied to observations at varying resolutions.

\begin{figure}[!b]
     \centering
     \vspace{-2mm}
     \includegraphics[width=11.2cm]{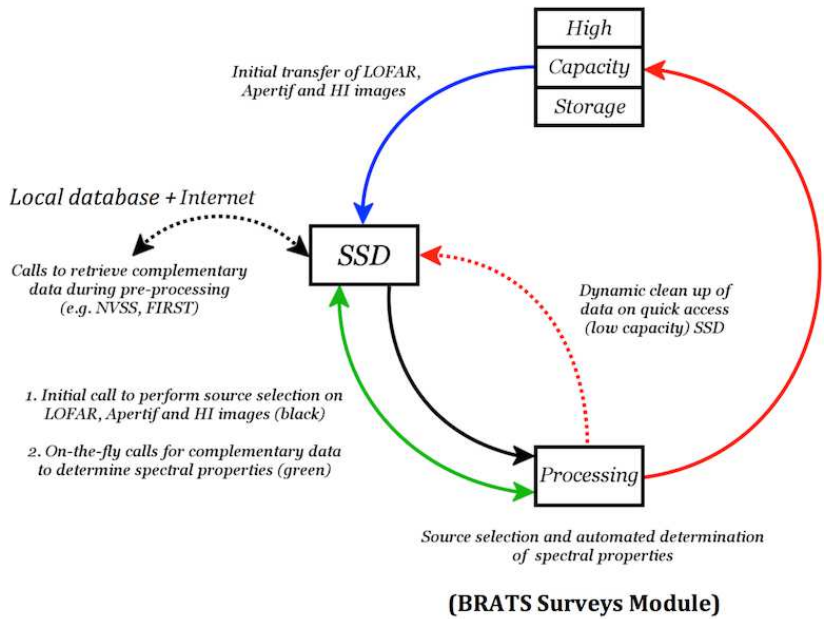}
     \caption{Preliminary process diagram for the analysis of LOFAR data using the \textsc{brats} survey module.}
     \label{processdiagram}
     \end{figure}

\section{The BRATS survey module}
\label{brats}

\textsc{brats} was originally designed as a user friendly way for determining the age of optically thin radio sources emitting via the synchrotron process through fitting of spectral ageing models \cite{kardashev62, pacholczyk70, jaffe73, tribble93, komissarov94} on small spatial scales \cite{harwood13, harwood15}. However, it has since expanded in to a much wider suite of analysis tools which, amongst other features, includes spectral index fitting, parameter determination, image manipulation and adaptive region selection. This existing infrastructure is therefore the ideal framework to build upon to efficiently identify and analyse interesting sources in large data volumes.

Fig. \ref{processdiagram} gives a basic overview of the process that will be applied by the \textsc{brats} surveys module, using processing of Low Frequency Array (LOFAR) wide-field images and surveys  as an example. Survey data are initially transferred from the long-term archive to a local processing machine (e.g. a local cluster) either in the form of a catalog, or as calibrated images depending on the type of analysis to be performed. At this stage, an optional initial cut may be applied to the data based on a source's basic properties such as flux density, position on the sky or whether a source is resolved. For each remaining source, the surveys module performs model fitting (spectral index, spectral age, or another project specific model) using the standard \textsc{brats} fitting procedures \cite{harwood13, harwood15}, making on-the-fly calls to cross-match with complementary archival data. For the initial release, this will comprise of a local catalog of complementary observations (e.g. NVSS and FIRST) but will later be expanded to search for these values automatically via Virtual Observatory (VO) services. Where complementary data are not available the source can either be flagged or, where possible, an upper limit used for the model fitting. Once fitting is complete, the results and any properties determined during the fitting process are then fed back in to the long-term archive (or other specified location) and the cycle repeated for the next source. The surveys module will be both multithreaded and multiprocess meaning that this cycle can be run simultaneously on multiple sources and fields on multi-core machines and distributed clusters, making it a highly scalable and time efficient analysis tool for a wide range of sample sizes and computing environments. It should be noted that while Fig. \ref{processdiagram} uses the processing of LOFAR data as an example, this same processing cycle will be applicable to any radio survey or sample.

While the process cycle described above is relatively straight forward to implement, one of the main hurdles which must be overcome in the selection and spectral analysis of sources in interferometric radio surveys is that they are usually only carried out at a single frequency and in a fixed array configuration. The advent of broad-bandwidth observations has meant that spectral index maps can be produced for a source within any single observation; however, these in-band spectra require a high signal to noise ratio and still only cover a relatively small range of frequency space. Uncertainties due to absolute flux calibration errors and the thermal noise therefore make in-band spectra alone unsuitable for source selection, particularly in cases where identifying possible curvature in a spectrum is required. These problems are further compounded by the fact that the varying resolution, UV coverage and sensitivity between surveys mean it is difficult to ensure that the complementary data are comparable to that of the new observations in an automated way. A multistage approach is therefore required if one is to reliably identify potentially interesting sources and maximise the level of analysis performed for a source given the available data. In the following sections we discuss how \textsc{brats} will approach these problems, ranging from unresolved sources and upper limits up to the detailed analysis of well resolved sources over a wide frequency range. While here we discuss this process in the context of established and remnant radio galaxies, \textsc{brats} is designed in such a way that any model (e.g. models with a thermal component, CSS/GPS models) can be easily added to this framework for use with a wide range of potential science projects.

\begin{figure}[!b]
     \centering
          \vspace{-2mm}
     \includegraphics[width=9.2cm]{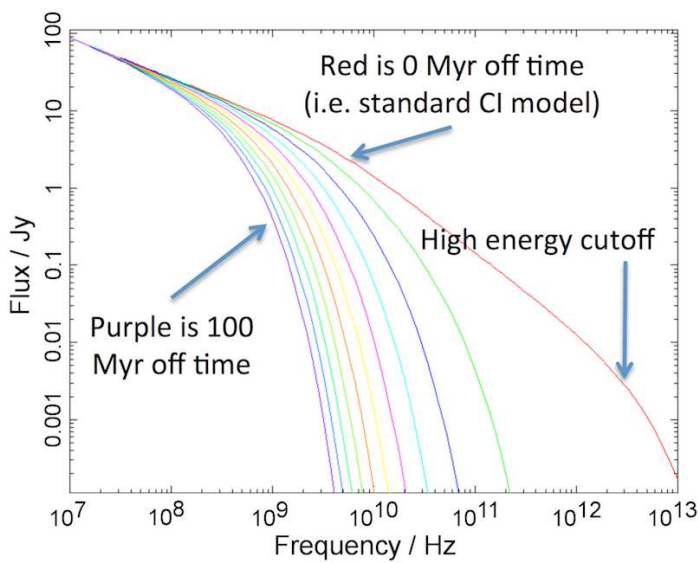}
     \caption{Example CI off (KGJP) model of spectral ageing \cite{komissarov94} with an off time between 0 and 100 Myr at 10 Myr intervals. The active phase is fixed at 30 Myr with an arbitrary normalisation. Note that the red line where $t_{off}=0$ is equivalent to the standard continuous injection model \cite{pacholczyk70}.}
     \label{cioff}
     \end{figure}

\subsection{Integrated flux}
\label{integrated}

The limited resolution of many surveys performed both historically (e.g. NVSS, WENSS) and more recently at low-frequencies (e.g. the LOFAR Multifrequency Snapshot Sky Survey, MSSS \cite{heald15}; the Galactic and Extragalactic MWA survey, GLEAM \cite{wayth15}), means that the majority of sources are at least partially, and in many cases fully, unresolved. Even for higher resolution surveys such as FIRST, the complementary data at a comparable resolution required to perform a resolved spectral analysis is often not available. For upcoming and future surveys (e.g. LOFAR tier-1 survey, Shimwell et al., in prep), and for individual targeted wide-field observations (e.g. the Bootes field \cite{williams16}), this problem is further compounded by the fact that they will be much more sensitive than current archival data and so in many cases upper limits, rather than detections, must be used. An initial selection and analysis based on the integrated flux of potential targets is therefore vital for any survey or large sample.

\begin{figure}[!b]
     \centering
          \vspace{-2mm}
     \includegraphics[width=14.5cm]{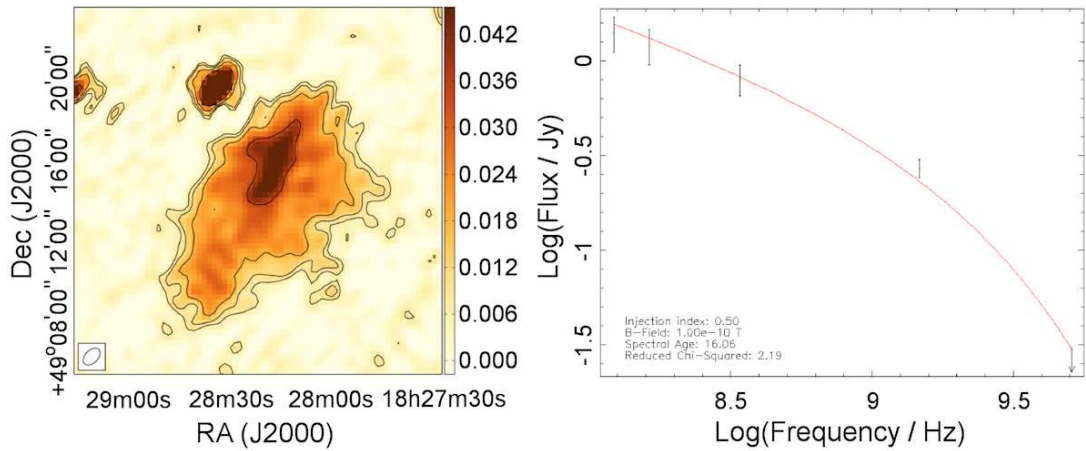}
     \caption{Left: 137 MHz LOFAR image of the remnant radio galaxy BLOB1 \cite{brienza16}. Right: \textsc{brats} fitting of the CI off model to BLOB1 using 116 and 155 MHz LOFAR, 327 MHz WENSS, 1.4 GHz WSRT, and 4.85 GHz GB6 (upper limit) data. The best fitting model provides a characteristic lower limit on the age of 76 Myr ($t_{on} = 16$ Myr, $t_{off} = 60$ Myr), in agreement with previous results \cite{brienza16}.}
     \label{blob1}
     \end{figure}

For radio galaxies, the standard models of spectral ageing used when dealing with integrated fluxes (Fig. \ref{cioff}) are the continuous injection (CI) model for active sources \cite{pacholczyk70} and the CI off model for remnant galaxies (also known as the KGJP model, \cite{komissarov94}), and have been successfully applied to small samples of galaxies in the past (e.g. \cite{carilli91, slee01, parma07}). Both of these models are available in \textsc{brats}\footnote{Version 2.5 onwards} and have been well tested against both historic and recent studies (e.g. Fig. \ref{blob1}). The main advantages of this form of fitting are that these models are computationally inexpensive (usually less than a few seconds per source), do not require the original radio maps to be available, and can also handle limited data quality such as upper limits, but this quick fitting and flexibility in data quality comes at a cost. Along side being unable to determine variations as a function of position, for large samples generic assumptions must be made about the model parameters, such as magnetic field strength and injection index (a parameter which describes the initial electron distribution of the source). Even if these parameters can be well constrained or approximated for a given sample, the ability of these models to reliably recover physical parameters such as its age has been called in to question. One of the underlying assumptions made for these models is that parameters such as magnetic field and jet power, remain approximately constant over the lifetime of the source; however, simulations of radio galaxies have suggest that this is unlikely to be the case \cite{kapinska15}, with properties such as the the low-frequency spectral index of the source varying non-linearly as a function of source length (Hardcastle et al., in prep).

Even with these draw backs, fitting of spectral ageing models to the integrated flux can still prove a powerful source selection tool. While they may not currently be able to determine a highly accurate age for a source, there remains a distinct difference between the spectrum of an active source described by the CI model, and a remnant source describe by the CI off model. Therefore if, for example, one wishes to search a sample or survey for remnant radio galaxies it provides an invaluable method of identifying potential sources. Through the \textsc{brats} survey module this form of integrated fitting, whether using models of spectral ageing, spectral index or models describing other source types, will therefore form the starting point for most survey searches, reducing the large number of observed sources down to a list of likely candidates in an automated manner that is suitable for a more detailed follow up analysis or, where a robust model and selection criteria exists, allow population studies to be undertaken. Determination of the required criteria for these type of systematic searches are already currently being undertaken for LOFAR fields in the search for remnant radio galaxies (see Brienza et al., these proceedings) which will soon be applied using \textsc{brats} on a variety of observed fields (e.g. the Lockman hole, Bootes fiedl) and subsequently to the LOFAR tier-1 survey.

\subsection{Resolved sources}
\label{resolved}

While selection based on an analysis of the integrated flux of a source forms a good starting point, the fitting of models on smaller spatial scales can provide much greater insight in to the underlying physics of individual sources and populations as a whole. The automation of such fitting is a non-trivial process requiring factors such as the availability of well aligned images matched in resolution and UV coverage to be considered. However, over the next few years the availability of high resolution, high image fidelity surveys are set to become more common place meaning that well matched, high resolution ancillary data will exist over a wide frequency range (e.g.  120-200 MHz: LOFAR tier-1; 1-2 GHz: MeerKAT MIGHTEE, APERTIF; 2-4 GHz: VLASS).

\begin{figure}[!b]
     \centering
          \vspace{-2mm}
     \includegraphics[height=5.8cm]{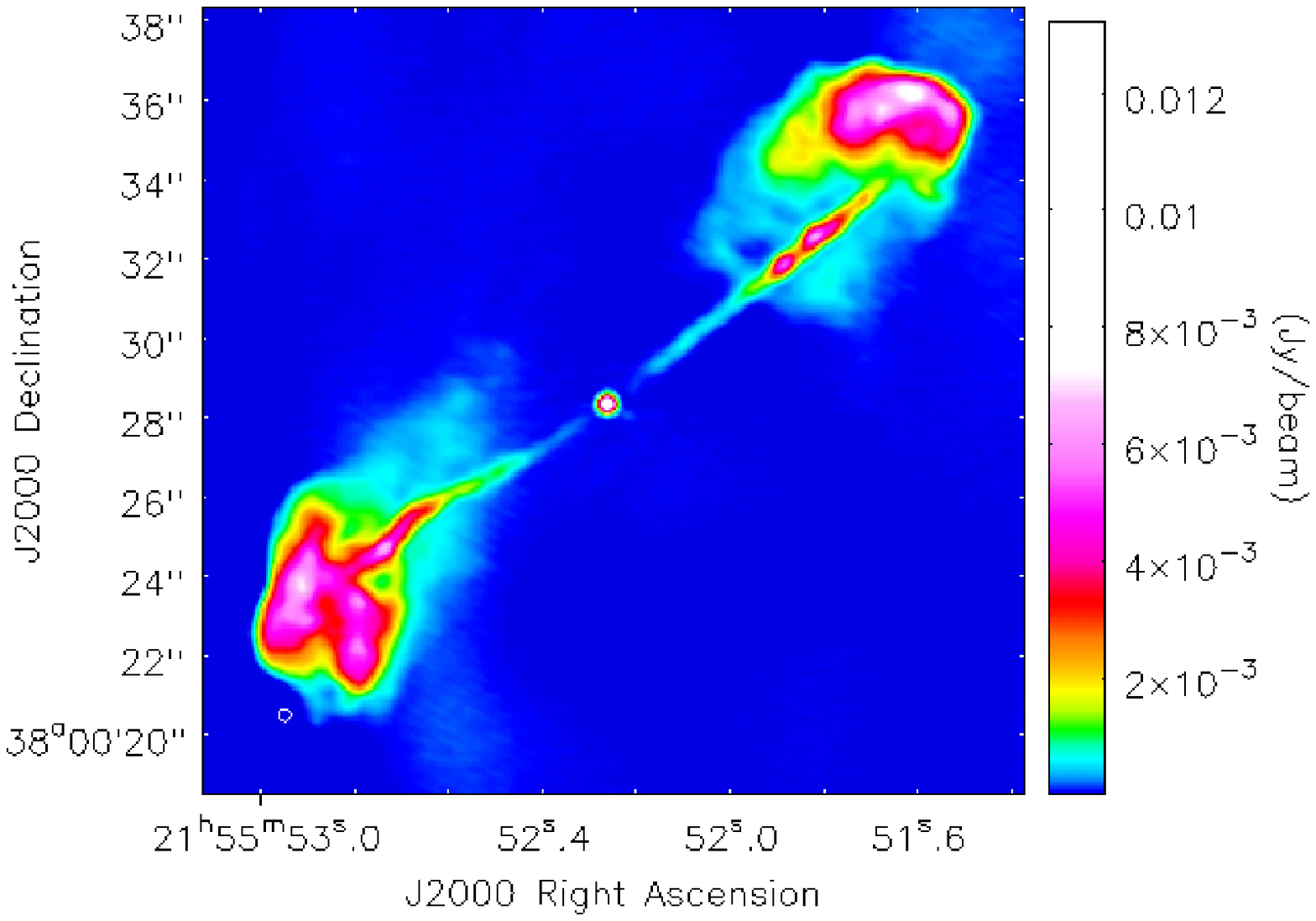}
     \includegraphics[height=5.8cm]{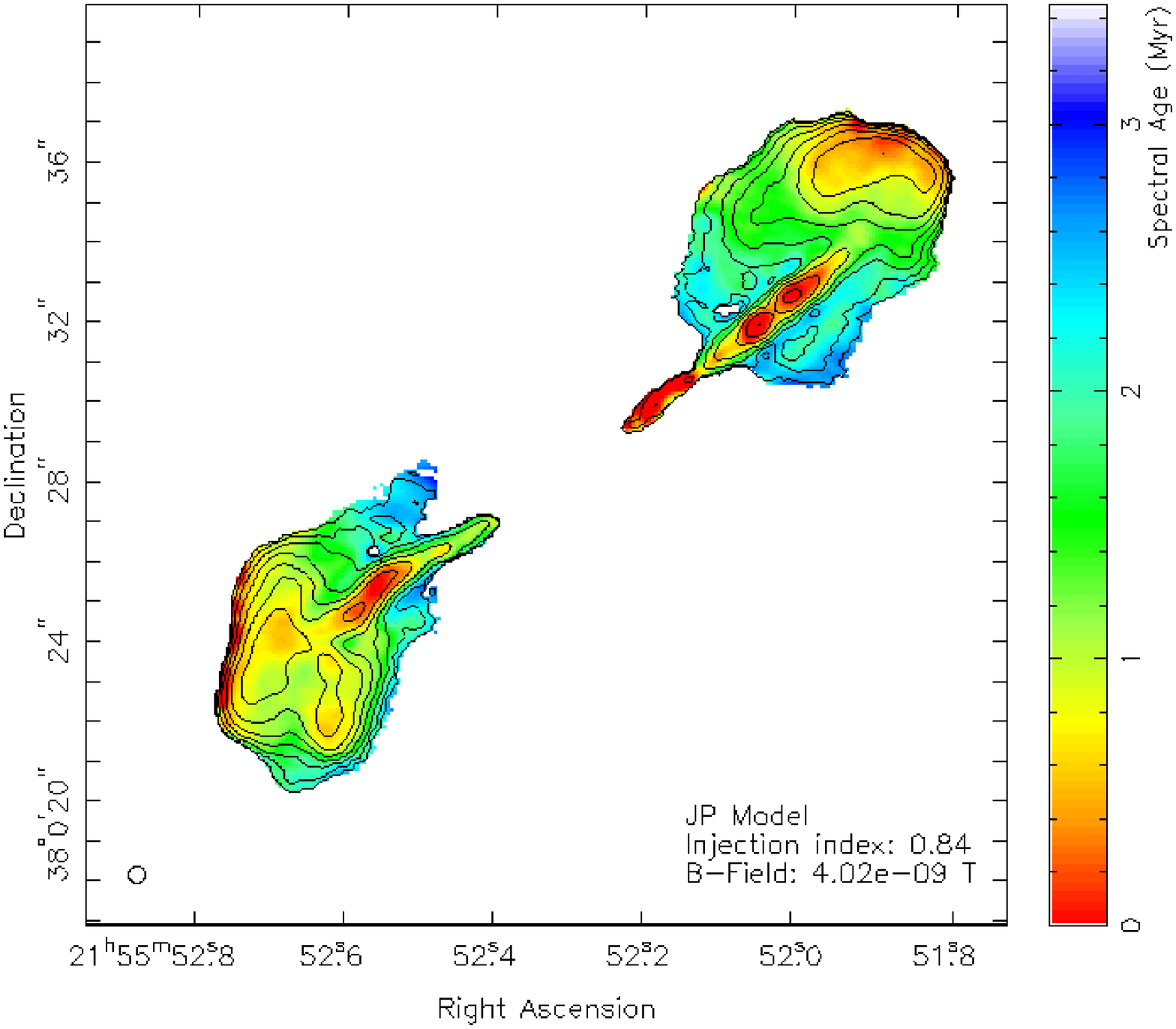}
     \caption{Left: Well resolved, 6 GHz VLA image of 3C438 using combined A, B and C-configurations with 4 GHz bandwidth. The beam size, shown in the bottom left corner, is 0.3 arcsec with an off-source RMS of 8 $\mu$Jy beam$^{-1}$. Right: Spectral age map of 3C438 on small spatial scales using new, broad-bandwidth 4 to 8 GHz VLA observations and archival 1.4 GHz data. The maximum age of the source for the JP model of spectral ageing (shown here) is 3 Myr \cite{harwood15}.}
     \label{3c438}
\end{figure}

Fitting of spectral ageing models, such as the JP and KP models \cite{kardashev62, pacholczyk70, jaffe73} on which the CI (off) models are based, to well resolved sources is a well established function of the \textsc{brats} software package but usually requires a manual alignment of the images on sub-pixel scales. Fortunately, the \textsc{brats} adaptive regions function, which automatically creates regions based on a set of user parameters such as signal to noise ratio, can help negate this problem. For any given source, the surveys module will allow regions to be allocated based on both the standard user parameters, and an assumed alignment error which can be determined based on checking only a small subset of the survey sample. Assuming the images are at least reasonably well aligned (i.e. an alignment error much smaller than the beam size), models can be fit without the need for the high, sub-pixel levels of accuracy. Such moderately resolved fitting will form an important part of the next generation of radio surveys allowing the identification and analysis of populations (e.g. recently switched off radio galaxies) that cannot be be selected based on the integrated flux alone.
     
The computational requirements needed to run this form of moderately resolved analysis are, unsurprisingly, greater than those for the integrated flux, but fitting of spectral ageing models in this way still only requires between 1 and 5 minutes per source depending on the number of fitted regions and the number of data points available\footnote{Here we assume a single, 12 core computing node}. Given current computing facilities, running such an analysis over an entire whole sky survey, although possible, is less than desirable (and in most cases, unnecessary). For most science cases, moderately resolved fitting will therefore form the second stage of selection an analysis using a sample initially selected through integrated fitting, or the first stage of individual, targeted wide-field observations.

Fitting of models on small, pixel by pixel, scales (an example of which is shown in Fig. \ref{3c438}) is also likely to play a major role in the analysis of survey data. In allowing variations as a function of position to be accurately determined and reducing the detrimental effects associated with fitting over large areas of a source such as the superposition of spectra \cite{harwood13, harwood15}, this small scale analysis can provide much deeper insight in to the underlying physical processes of a source. Details of model fitting on very small scales is discussed at length in papers by Harwood et al. \cite{harwood13, harwood15} and the \textsc{brats} user manual, and so we do not repeat it here; however, we note that for the foreseeable future this step will require manual intervention for the alignment of images on the sub-pixel scales and significant computational overheads. This type of highly detailed study is therefore likely to form the final stage of a survey in order to further explore the underlying physics in a handful of interesting or archetypal sources for which targeted follow up observations have been made over a large frequency range.

\section{Conclusions}
\label{conclusion}

In these proceedings we have presented an overview of the \textsc{brats} surveys module and a pathway using these techniques from an initial selection process of unresolved sources through to a highly detailed analysis on small scales for recent and upcoming radio surveys and samples. The key points made are as follows:

\begin{enumerate}
\item The large number of potentially interesting sources found in current and future surveys requires an automated approach to source selection and analysis.
\item The \textsc{brats} surveys module will provide tools allowing the selection and analysis of sources based on their spectral properties on relatively short timescales.
\item Such automation is required if one is to fully explore the properties of any given population as a whole.
\item \textsc{brats} and the surveys module is designed such that new and additional models can be easily included to maximise the range of possible science goals.
\item Optimising the science obtained from surveys and large samples will require a multi-stage approach; progressing from large to small scales.

\end{enumerate}

Model fitting and region selection of individual sources using \textsc{brats} is now a well established process \cite{harwood13, heesen14, stroe14, harwood15, degasperin15}. Phase 1 of the surveys module will provide a method of automatically performing model fitting from a local database, with phase 2 expanding this capability to include the automated retrieval of complementary data using VO services. While this will likely prove invaluable in the short to medium term for the selection and analysis of sources in ongoing and planned surveys, we foresee development of these techniques continuing based on the needs of users well into the SKA era.

\section{Acknowledgements}
\label{acknowledgements}
We wish to thank Marisa Brienza for her helpful comments on these proceedings. The research leading to these results has received funding from the European Research Council under the European Union's Seventh Framework Programme (FP/2007-2013) / ERC Advanced Grant RADIOLIFE-320745.

\end{document}